\documentclass{iopjournal}

\usepackage{amsmath,amssymb}
\usepackage{bm}
\usepackage{dcolumn}
\usepackage{float}
\usepackage[export]{adjustbox}
\newcommand{\revred}[1]{\textcolor{black}{#1}}
\newcommand{\revredd}[1]{\textcolor{black}{#1}}

\let\OriginalIncludeGraphics\includegraphics
\renewcommand{\includegraphics}[2][]{%
  \OriginalIncludeGraphics[#1,max width=\linewidth,max totalheight=0.72\textheight,keepaspectratio]{#2}%
}

\makeatletter
\renewenvironment{figure*}{\@float{figure}}{\end@float}
\makeatother

\newcommand{\KAISTQST}{Graduate School of Quantum Science and Technology, Korea Advanced Institute of Science and Technology, 291 Daehak-ro, Yuseong-gu, Daejeon 34141, Republic of Korea}
\newcommand{\KRISS}{Korea Research Institute of Standards and Science, 267 Gajeong-ro, Daejeon 34113, Republic of Korea}
\newcommand{\NNFC}{Center for Next-Generation Platform Development, National NanoFab Center (NNFC), Daejeon 34141, Republic of Korea}

\begin{document}

\articletype{Paper}

\title{Strontium ${}^{1}S_{0}\!\rightarrow\!{}^{1}P_{1}$ transition frequency measurements assisted by a photonic grating chip}

\author{Jaewhan Lee$^{1,2,\dagger}$, Hyun Gyung Lee$^{1,\dagger}$, Won-Kyu Lee$^1$, Huidong Kim$^1$, Dohyeon Kwon$^1$, Sang-Bum Lee$^1$, Meungho Seo$^1$, Taeg Yong Kwon$^1$, Sangwon Seo$^1$, Hyun-Gue Hong$^1$, Seji Kang$^1$, Sang Eon Park$^1$, Young-Ho Park$^1$, Jongcheol Park$^3$, Yeeun Na$^3$, Il-Suk Kang$^3$, Sangsik Kim$^2$ and Jae Hoon Lee$^{1,2,*}$}

\affil{$^1$\KRISS}\par
\affil{$^2$\KAISTQST}\par
\affil{$^3$\NNFC}\par
\affil{$^\dagger$These authors contributed equally to this work.}\par
\affil{$^*$Author to whom any correspondence should be addressed.}\par

\email{jhloptics@kriss.re.kr}

\keywords{strontium, absolute frequency, fluorescence spectroscopy, grating magneto-optical trap, photonic grating chip, wavelength-meter calibration}

\begin{abstract}
We measure the absolute frequency of the ${}^{1}S_{0}\!\rightarrow\!{}^{1}P_{1}$ transition in strontium using two methods: fluorescence spectroscopy of a thermal atomic beam source from a compact low-power oven and velocity measurements of a slow atomic beam from a two-dimensional grating magneto-optical trap (2D gMOT). 
The measurements for both methods are performed in the same ultra-high vacuum chamber containing a diffraction grating chip which is placed below the strontium atoms that are being interrogated.
The first method uses a probe laser beam incident on the grating chip such that the grating acts as an end mirror, with the first-order diffracted beam providing a retro-reflected probe beam.
The counter-propagating laser beams traverse an atomic beam emitted from an oven, enabling spatially resolved fluorescence spectroscopy through CCD imaging and hyperfine-constrained multi-isotope fitting.
The second method relies on a large profile cooling laser beam normally incident onto the grating chip which laser cools strontium atoms for a slow atomic beam source. 
The velocity of the atoms exiting the 2D gMOT is measured as a function of the laser detuning and intensity from which the resonance frequency can be estimated. 
The two methods are consistent within their quoted uncertainties. 
Using three datasets based on retro-beam spectroscopy measurements, and one dataset using slow atom beam velocity measurements, we determine the ${}^{1}S_{0}\!\rightarrow\!{}^{1}P_{1}$ transition frequency to be $650.503\,815(5)~\mathrm{THz}$. Our result provides a re-evaluation of this $461$ nm transition demonstrated on a compact laser cooling apparatus based on a diffraction grating platform.
\end{abstract}

\section{Introduction}
The ${}^{1}S_{0}$ ground state to ${}^{1}P_{1}$ excited state transition is a foundational component of strontium-based quantum systems through its central role in laser cooling and trapping\cite{courtillot2003,xu2003}, fluorescence detection, and precision and isotope-selective spectroscopy\cite{Courtillot2005,poli2006,nagel2005,bushaw2000,miyake2019}.
The development of next-generation Sr quantum systems demands more stringent accuracy in the $461~\mathrm{nm}$ absolute wavelength for applications ranging from optical lattice clock evaluation\cite{ludlow2015,hobson2020,targat2013,zheng2022,oelker2019}, magic-wavelength estimations\cite{Safronova2013,Kestler2022,Ma2025,Kestler2026}, atomic-beam slowing and cold atomic beam generation\cite{kwon2023,nosske2017}, the design of chip-based cold-atom platforms\cite{lee2025,pate2023,sitaram2020,Nshii2013,Imhof2017,bondza2022}, and quantum computing\cite{heinz2020}.

\revred{Currently, the widely referenced experimental determination of the ${}^{88}\text{Sr}$ ${}^{1}S_{0}\rightarrow{}^{1}P_{1}$ transition frequency originates from a measurement performed nearly nine decades ago~\cite{sullivan1938}. The `Observed value' listed in the NIST Atomic Spectra Database (ASD)\cite{NIST_ASD}, which cites this 1938 work, is 650.503\,47(31)~THz. 
	Compilations of Sr spectroscopic data~\cite{sansonetti2010} often report the `Ritz value' rather than this observed measurement, which can lead to discrepancies in the literature. 
	Subsequent work has continued to rely on these older values, together with updated isotope shifts, to infer transition frequencies for ${}^{86}\text{Sr}$ and ${}^{87}\text{Sr}$.}

In this work, we implement a photonic grating chip designed for $461~\mathrm{nm}$ that generates an optimal diffraction geometry necessary for laser cooling from a circularly polarized input\cite{lee2025}. The grating chip also serves as a stable platform for generating the counter-propagating probe beams used for fluorescence spectroscopy\cite{Isichenko2023,Hummon2021}. This in-vacuum, chip-integrated approach replaces traditional external mirrors, significantly enhancing the mechanical stability of the apparatus for precision metrology.

\begin{figure*}[htbp]
	\centering
	\includegraphics[width=0.55\columnwidth]{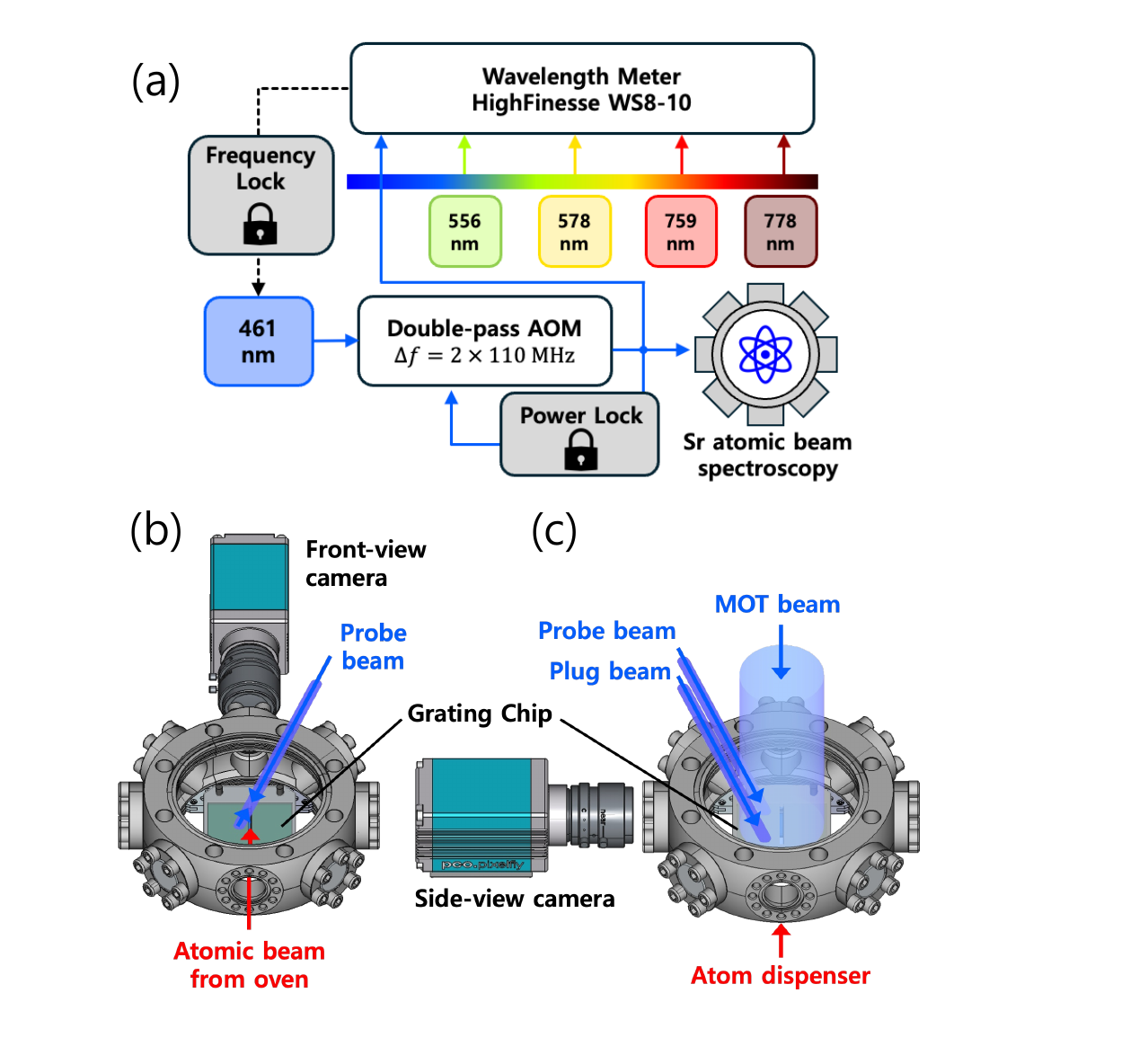}
	\caption{
		(a) Schematic of the laser frequency locking system used for the $461~\mathrm{nm}$ spectroscopy beam.
		A $461~\mathrm{nm}$ laser is frequency shifted by a double-pass AOM before being delivered to the vacuum chamber for interacting with the strontium atoms.
		Auxiliary lasers at 556, 578, 759, and 778 nm are monitored by a wavelength meter (HighFinesse WS8-10), while the $461~\mathrm{nm}$ laser is frequency and power stabilized by PID feedback loops.
		Experimental apparatus showing the two types of chip-defined probe geometries for (b) the retro-beam spectroscopic measurements and (c) the 2D gMOT laser-cooled atom velocity measurements.
		The $461~\mathrm{nm}$ probe and plug beams enter the chamber obliquely to intersect the
		atoms. Fluorescence is collected by a front-view or side-view imaging system.
	}
	\label{fig:apparatus_schematic}
\end{figure*}

Fig.~\ref{fig:apparatus_schematic} shows a schematic of the frequency-locked laser system and the experimental apparatus. For spectroscopic measurements, fluorescence from thermal atoms interacting with counter-propagating laser beams after emerging from an oven is recorded with a CCD camera above the grating chip across a scan of probe frequencies. A three-dimensional tensor $\mathcal{S}(x,y,f)$ is produced from the image set, in which each pixel contains a complete spectrum of the ${}^{1}S_{0}\!\rightarrow{}^{1}P_{1}$ transition. Fitting each spectrum with a hyperfine-constrained multi-isotope pseudo-Voigt model yields spatially resolved maps of the line center, linewidth, and isotope contributions across the atom-light interaction region.

Complementing the spectroscopic approach, we also perform an evaluation of the resonance frequency based on the kinematic properties of the slow atomic beam generated by a 2D gMOT. Unlike thermal atomic beams, the axial velocity of atoms exiting the 2D gMOT is intrinsically determined by the transverse capture velocity, which is governed by the cooling laser parameters\cite{nosske2017,lee2025,Dieckmann1998}. By precisely mapping the velocity distribution across a range of laser conditions and applying an elementary model based on the photon scattering rate, the atomic resonance frequency can be extracted as a free parameter.

Using these two methods, we establish three retro-beam spectroscopy estimates together with one 2D gMOT-based estimate for the ${}^{88}$Sr ${}^{1}S_{0}\!\rightarrow\!{}^{1}P_{1}$ transition frequency. Combining these determinations yields $650.503\,815(5)~\mathrm{THz}$. To our knowledge, this work constitutes the first reported laboratory re-determination of this transition frequency since 1938, offering a substantial refinement in both absolute accuracy and precision.

\begin{figure*}[htbp]
	\centering
	\includegraphics[width=0.9\columnwidth]{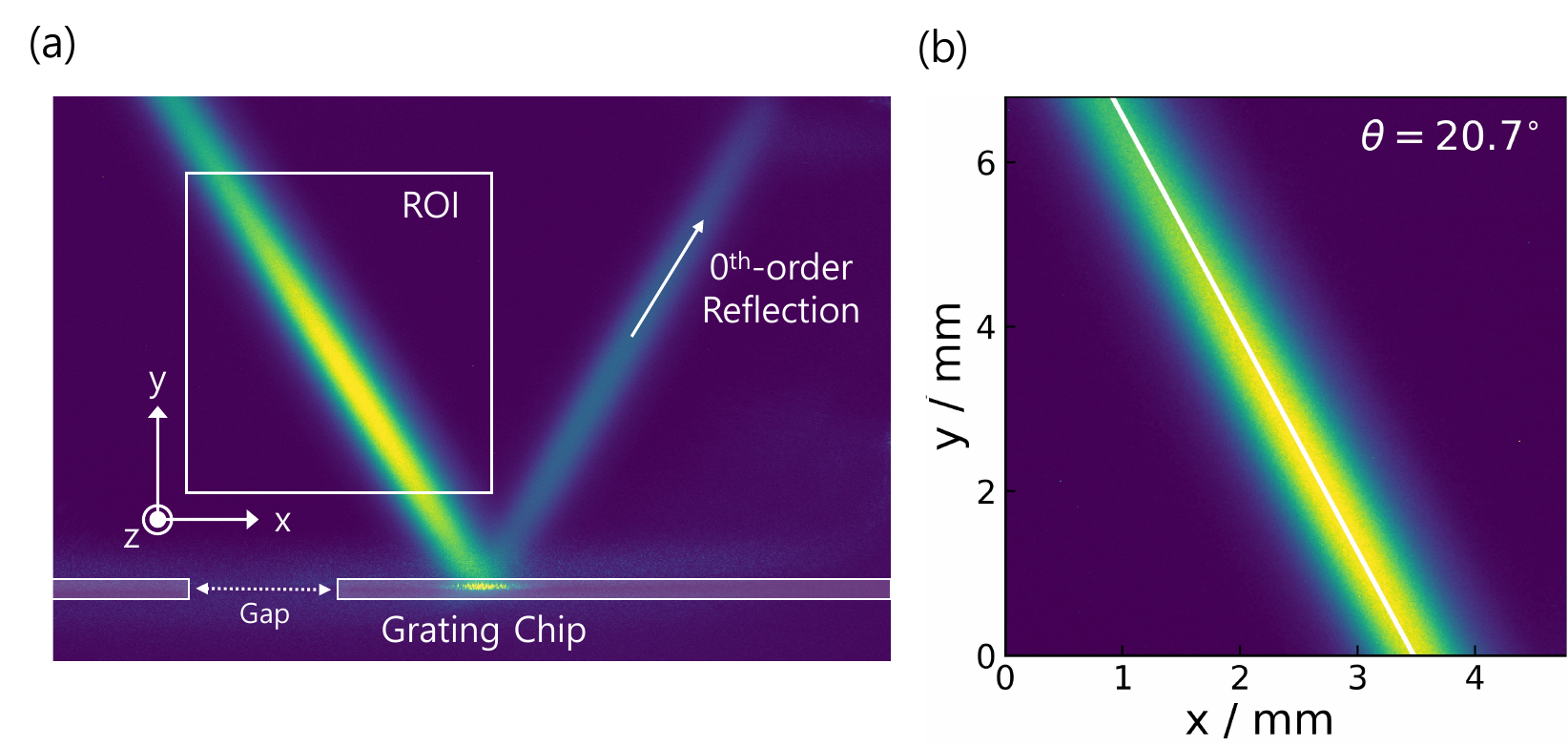}
	\caption{
		(a) CCD fluorescence image recorded for the $461~\mathrm{nm}$ probe beam above the chip.
		The bright elongated trace marks the probe--atom interaction region formed by
		the intersection of the probe beam with
		the thermal strontium atomic beam.
		The white box marks the region of interest
		(ROI) selected for beam-angle analysis.
		(b) Enlarged view of the ROI.
		A principal-component analysis (PCA) of the
		fluorescence distribution yields the first principal axis (white line),
		giving an in-plane incidence angle of $\theta = 20.70^\circ$.
	}
	\label{fig:probe_geometry}
\end{figure*}

\section{Methods}
We employ two complementary methods to characterize the strontium ${}^{1}S_{0}\!\rightarrow\!{}^{1}P_{1}$ transition resonance. 
The experimental campaign yielded four independent datasets, categorized by technique for clarity: R1, R2, and R3 denote trials conducted via retro-beam fluorescence spectroscopy, while V1 represents the measurement derived from the 2D gMOT velocity mapping.

\revred{The in-vacuum grating chip, which forms the foundation for both experimental methods, was implemented as an Al-coated silicon reflective diffraction platform. The fabrication process was based on our previously demonstrated wafer-scale grating-chip platform~\cite{lee2025}. The grating structures were fabricated on 200~mm silicon wafers using CMOS-compatible 193~nm deep ultraviolet lithography and plasma etching. The wafer-level design linewidth was set to 229~nm with a spacing of 423~nm to compensate for the linewidth increase induced by the subsequent metal coating. After silicon etching, a thin Al-based reflective metal layer was deposited by sputtering to form the final reflective diffraction surface. The final Al-coated grating was designed to have a ridge height of 120~nm, a linewidth of 249~nm, and a period of 652~nm. The fabricated grating chip was diced into $20~\text{mm} \times 30~\text{mm}$ chips from the 200~mm wafer. A representative transmission electron microscopy (TEM) cross-section showed a ridge height, linewidth, and period of 119.2~nm, 247.7~nm, and 652.0~nm, respectively, corresponding to less than 1\% deviation from the target ridge geometry. The across-wafer critical dimension (CD) uniformity was 0.61\%. The measured first-order diffraction angle and diffraction efficiency for a normally incident 461~nm cooling beam were $45.0^\circ$ and 28.5\%, respectively, and the measured s--p phase shift was $23.9^\circ$. These characteristics provide the diffracted beam geometry applicable for both the spatial fluorescence spectroscopy and the 2D gMOT operation.
}

\subsection{Spatially resolved atomic fluorescence spectroscopy}
\label{sec:spectroscopyMethod}

The first method implemented to determine the strontium ${}^{1}S_{0}\!\rightarrow\!{}^{1}P_{1}$ resonance employs image-based fluorescence spectroscopy.  
The in-vacuum photonic grating chip establishes a rigid counter-propagating beam geometry, facilitating precise measurements of first-order Doppler shifts. 
A CCD camera acquires two-dimensional images, such as that shown in Fig.~\ref{fig:probe_geometry}, for various laser frequencies to obtain spatially resolved fluorescence spectra.
Pixel-wise multi-isotope fitting yields spatial maps of the Doppler-free resonance value
that are statistically averaged for all relevant pixels to obtain dataset-level resonance frequency estimates. 

\revred{
	Compared to conventional external or in-vacuum mirror-based retro-reflection setups, the proposed in-vacuum grating chip architecture offers several distinct practical advantages. First, the distance from the atoms to the retro-reflection point is highly minimized to approximately 5~mm. This close proximity strongly suppresses errors and noise in the laser intensity caused by retro-reflection angle vibrations. While an in-vacuum mirror could achieve similar proximity, the grating chip is already integrated for the 2D gMOT, allowing it to be readily utilized without introducing additional components. Second, because the diffraction efficiency is discernibly lower than unity ($\sim$48\%), the two spectral peaks corresponding to the counter-propagating beams for each isotope exhibit a noticeable difference in peak height. This asymmetry makes the fitting of the spectra unambiguous in terms of identifying the up-shifted and down-shifted Doppler peaks, allowing the direction of the Doppler shift to be easily obtained. Third, the period of the diffraction pattern on the grating chip is known to better than 0.1~nm, as confirmed via TEM imaging and laser diffraction angle measurements. This ensures a highly precise retro-reflecting angle of 20.697(2)$^\circ$, which is utilized to accurately calibrate the $x$ and $y$ dimension ratios for the CCD pixels. 
	A potential disadvantage of utilizing a diffraction grating rather than a standard mirror is that the retro-reflection angle is intrinsically wavelength-dependent. However, because the wavelength scanning range for this experiment is only $\sim$500~MHz, the corresponding retro-reflected beam angle deviation is only $2 \times 10^{-5}$ degrees. This beam angle uncertainty is negligible in the current study, as it is approximately 3,000 times smaller than the geometrical angle uncertainty arising from standard experimental beam alignment, which is evaluated in section~\ref{sec:GeomAlignUncert}. Details of this method are described below.
}

\subsubsection{Data acquisition and selection}
\label{data}

In thermal-beam spectroscopy, the probe beam direction is typically configured to be perpendicular with respect to the atomic trajectory from an oven when trying to suppress Doppler shifts. This technique is extremely sensitive to angular alignment; even a $1^\circ$ deviation can induce Doppler shifts on the order of several \revredd{megahertz}~\cite{Laup2020Yb}. In order to accurately extract this systematic uncertainty, we implement a spatially resolved fluorescence imaging technique. 
A collimated $461~\mathrm{nm}$ probe beam with a $1/e^2$ diameter of $1.32~\mathrm{mm}$ illuminates the photonic grating chip at the incidence angle that corresponds to an identical first-diffracted beam angle.
This in-vacuum chip-defined retro-reflection enables the counter-propagating beams to be passively stable with high overlap. 

Figures~\ref{fig:probe_geometry}(a) and (b) illustrate the probe-beam trajectory as a fluorescence track above the chip. 
\revred{The fluorescence image is captured with a CCD camera (PCO Pixelfly USB) and zoom lens (Navitar MVL7000) along the direction of the atomic beam, as shown in Fig.~\ref{fig:apparatus_schematic}(b).
}
To characterize the geometry, we employ a principal-component analysis of the intensity distribution, which yields the intensity-weighted centroid and the first principal axis (indicated by the white line). This axis defines the beam path, corresponding to an in-plane incidence angle of $\theta = 20.7^\circ$.
The incident probe beam is confined to the $xy$-plane, while the strontium atoms—originating from an oven nozzle $350~\mathrm{mm}$ upstream—propagate primarily along the $z$-direction. The observed Doppler splitting originates from the opposing in-plane wavevectors, $(\pm k_x, \pm k_y)$, of the incident and first-order diffracted beams. Consequently, atoms with non-zero transverse velocity components experience Doppler shifts of opposite sign. This splitting varies spatially according to the local velocity projection onto the wavevectors. The characteristic frequency-shifted dual-peak structure of the obtained spectra are used for the subsequent analysis.

The atomic fluorescence above the chip is recorded during a frequency scan of the probe laser, yielding a three-dimensional data tensor comprised of a stack of 2D images. \revred{To minimize the effects of irregularities between camera pixel values, two subsequent images are acquired for each laser frequency: one with the atomic beam present and a reference image without it. The final image used for the analysis is the background-subtracted two-dimensional image utilizing a 14-bit dynamic range. The camera gain is carefully optimized such that the dynamic range is adequately utilized while maintaining a linear response without saturating any pixels.} To suppress high-spatial-frequency CCD noise, the raw images undergo $5 \times 5$ spatial binning prior to fitting. This preprocessing stage enhances the signal-to-noise ratio while maintaining sufficient spatial resolution to characterize the millimeter-scale variations of the beam profile required for the spectroscopic analysis. In this representation, each downsampled pixel $(i,j)$ in the $xy$-plane corresponds to a complete ${}^{1}S_{0}\!\rightarrow\!{}^{1}P_{1}$ fluorescence spectrum, denoted as $\mathcal{S}_{i,j}(f)$.

To ensure the integrity of the final frequency determination, we apply two quality-based filtering criteria to the data tensor. First, pixels with insufficient signal-to-noise ratios are excluded using an intensity threshold set at $0.001\%$ of the global peak fluorescence. Second, we restrict the analysis to the spatial footprint of the probe beam by retaining only those pixels within a fixed distance from the beam axis, as defined by a Gaussian fit to the fluorescence profile. This dual-masking approach ensures that only well-defined, signal-bearing regions contribute to the spectroscopy analysis.

\subsubsection{Spatially resolved multi-isotope spectral fitting}
\label{fitting}

Before fitting the set of one-dimensional spectra $\mathcal{S}_{i,j}(f)$, each pixel spectrum is baseline-corrected by subtracting a first-order polynomial background offset. 
Each pixel spectrum is then fitted independently with a hyperfine-constrained multi-isotope pseudo-Voigt model containing contributions from ${}^{88}$Sr, ${}^{86}$Sr, and ${}^{87}$Sr isotopes as follows,
{
	\begin{equation*}
		\begin{aligned}
			\mathcal{S}(f)=
			&~A_{88}\Bigl[\alpha\,\mathcal{V}(f;f_{88}-\Delta_{\mathrm{D}},\Gamma,\eta)
			\\
			&\qquad\qquad
			+(1-\alpha)\mathcal{V}(f;f_{88}+\Delta_{\mathrm{D}},\Gamma,\eta)\Bigr]
			\\
			&+A_{86}\Bigl[\alpha\,\mathcal{V}(f;f_{86}-\Delta_{\mathrm{D}},\Gamma,\eta)
			\\
			&\qquad\qquad
			+(1-\alpha)\mathcal{V}(f;f_{86}+\Delta_{\mathrm{D}},\Gamma,\eta)\Bigr]
			\\
			&+A_{87}\sum_F w_F
			\Bigl[\alpha\,\mathcal{V}(f;f_{87,F}-\Delta_{\mathrm{D}},\Gamma,\eta)
			\\
			&\qquad\qquad
			+(1-\alpha)\mathcal{V}(f;f_{87,F}+\Delta_{\mathrm{D}},\Gamma,\eta)\Bigr].
		\end{aligned}
	\end{equation*}
}

Here, $\mathcal{V}(f; f_0, \Gamma, \eta)$ denotes a unit-area pseudo-Voigt profile centered at $f_0$ with a full-width at half-maximum (FWHM) of $\Gamma$ and a Lorentzian mixing parameter $\eta$,
defined as
{
	\[
	\mathcal{V}(f; f_0, \Gamma, \eta)=
	\eta\,\mathcal{L}(f; f_0, \Gamma)
	+(1-\eta)\,\mathcal{G}(f; f_0, \Gamma),
	\]
}
where $\mathcal{L}$ and $\mathcal{G}$ are unit-area Lorentzian and Gaussian profiles with the same FWHM.
Due to the extensive pixelwise dataset, we adopted a pseudo-Voigt profile as a computationally efficient approximation of the Voigt function. This choice significantly reduces the processing overhead without discernibly compromising the precision of the extracted resonance parameters, which is taken to account when evaluating the uncertainty.
The parameter $\Delta_{\text{D}}$ is the Doppler shift associated with a single probe beam; consequently, the separation between the counter-propagating resonances is $2\Delta_{\text{D}}$.
The asymmetry parameter $\alpha$ accounts for the area (intensity) difference between the Doppler-split peaks and is assumed to be common for all isotopic components.
$A_{88}$, $A_{86}$, and $A_{87}$ represent the total integrated areas for each respective isotope.
The ${}^{86}\text{Sr}$ contribution is defined by a single isotope shift relative to ${}^{88}\text{Sr}$, while the ${}^{87}\text{Sr}$ contribution is modeled using three fixed hyperfine offsets, $f_{87,F} = f_{88} + \Delta_{87,F}$, indexed by the total angular momentum $F$.
The weights $w_F$ are non-negative and satisfy the normalization $\sum_F w_F = 1$.
For each dataset, the pseudo-Voigt mixing parameter $\eta$ is selected from a stratified subset scan and then held fixed during the full tensor fit.

Global linear parameters---including the overall amplitude, constant baseline, and residual linear background---are handled efficiently via the \revred{variable projection (VarPro) algorithm\revred{\cite{GolubPereyra1973, OLearyRust2013}}.}
Finally, because the residual magnetic field in the interaction region is sufficiently small, any Zeeman splitting remains unresolved within the observed linewidth and signal-to-noise ratio. Therefore, the Zeeman term is not explicitly included in the fit model, but is considered for the uncertainty budget.

Reported values for the \revred{${}^{1}S_{0}\!\rightarrow\!{}^{1}P_{1}$} isotope shifts exhibit slight variations across different sources~\cite{lee2025, courtillotThesis, stellmerThesis, elgeeThesis}. Consequently, we treat the ${}^{86}\text{Sr}$ isotope shift, $\Delta_{86}$, as a constrained fit parameter within a literature-informed interval around $-125~\text{MHz}$.
For ${}^{87}\text{Sr}$, rather than employing a single effective isotope-shift parameter, we explicitly model the three hyperfine components relative to ${}^{88}\text{Sr}$ at $-9.7$, $-51.6$, and $-68.9~\text{MHz}$. These offsets correspond to the $F = 9/2 \to F' = 7/2, 11/2, \text{ and } 9/2$ transitions, respectively~\cite{stellmerThesis}. 

\begin{figure}[!t]
	\centering
	\includegraphics[width=0.95\columnwidth]{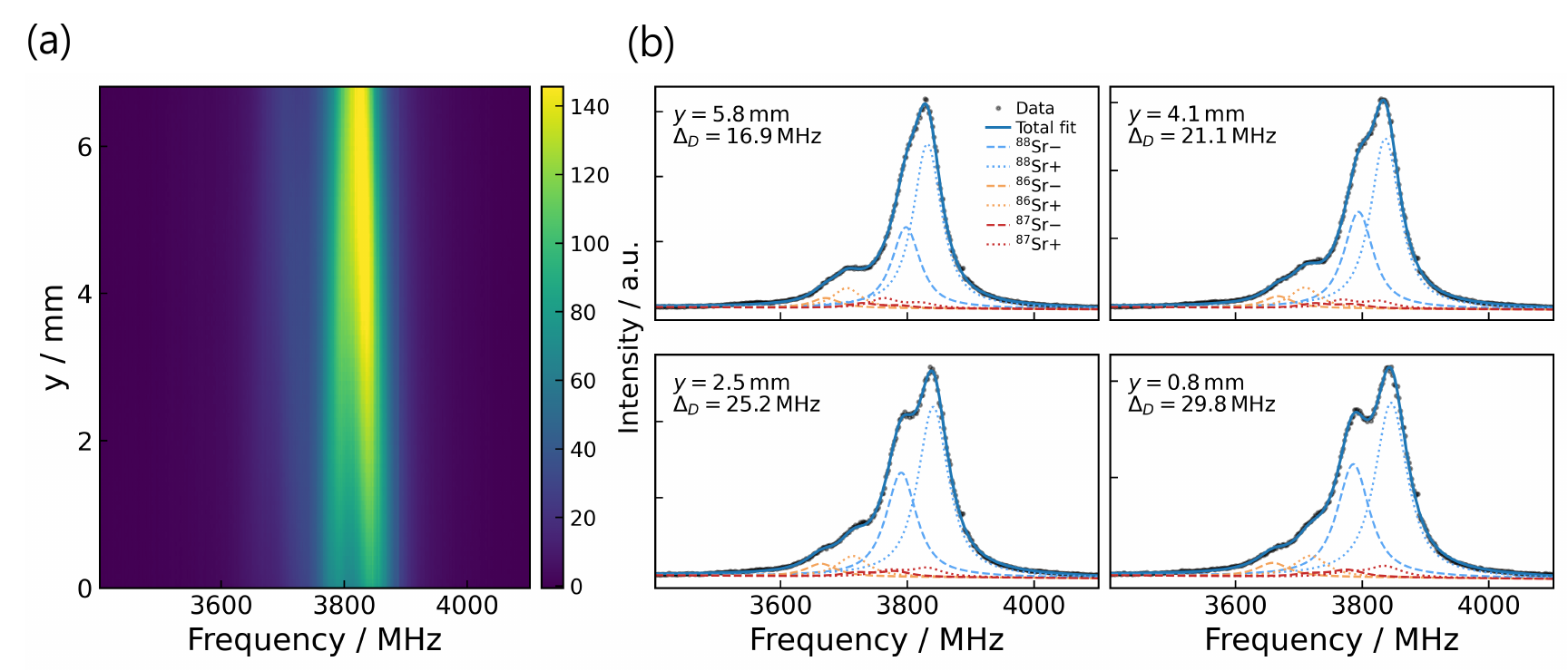}
	\caption{
		(a) Height--frequency fluorescence map measured above the grating chip.
		\revred{The frequency axis is plotted relative to $f_{\mathrm{base}}=650\,500\,000~\mathrm{MHz}$.
			The map is constructed in the PCA basis of the fluorescence track, with the
			horizontal direction following the probe-beam track and the vertical coordinate
			giving the height above the chip.
			(b) Representative spectra at four heights,
			$y=5.8$, $4.1$, $2.5$, and $0.8~\mathrm{mm}$,
			together with simultaneous multi-isotope pseudo-Voigt fits.
			The fitted single-beam Doppler shifts are
			$\Delta_{\mathrm{D}}=16.9$, $21.1$, $25.2$, and $29.8~\mathrm{MHz}$,
			respectively.} 
		Thin curves show the individual isotope and Doppler-branch
		components.
	}
	
	\label{fig:height_frequency_panel}
\end{figure}

Fig.~\ref{fig:height_frequency_panel} shows examples of the obtained fluorescence spectra and the fitted multi-isotope pseudo-Voigt functions for various positions along the $y$-direction from dataset R1. 
\revred{
	For Fig.~\ref{fig:height_frequency_panel}(a), the image shows the fluorescence along the principal axis defined by PCA of Fig.~\ref{fig:probe_geometry}. The horizontal coordinate follows the probe-beam frequency range, while the vertical coordinate gives the height above the chip. For each height bin, we
	average the spectra of the accepted pixels across the probe-beam ROI along the track direction, producing the height--frequency map. Note that the frequency axes are plotted relative to a baseline frequency
	$f_{\mathrm{base}}=650\,500\,000~\mathrm{MHz}$.
}
As clearly shown in Fig.~\ref{fig:height_frequency_panel}(b) the model successfully extracts the local resonance frequencies of the Doppler-split components for each isotope. A dependence of the Doppler shift to the $y$-axis (direction opposite to gravity) is observed due to the geometry of the probe region and the oven nozzle, which will be further discussed below.

\subsubsection{Spatial mapping and statistical uncertainty}
\label{selection}

\revred{After the fit is performed for each accepted pixel spectrum $\mathcal{S}_{i,j}(f)$, we calibrate the image dimensions to map the discrete pixel indices ($i, j$) to physical coordinates ($x, y$). This produces spatially resolved maps of the Doppler-free resonance frequency $f_{88}(x,y)$ and the Doppler shift $\Delta_{D}(x,y)$ across the interaction region, as shown in Fig.~4(a) and (c). Camera magnification, spatial resolution, and optical distortions primarily affect the assignment of spatial coordinates for these maps. The $x$ and $y$ axis scalings are calibrated via the imaging of a known grid pattern and subsequently fine-tuned using the known grating diffraction angle of 20.7$^\circ$ mentioned previously. We measure the distortion of the optical system to be approximately 0.9\% across the field of view. Because the spatial position of the CCD camera pixels does not define the laser frequency axis---that is, each individual pixel independently records a complete fluorescence spectrum over the probe-laser frequency scale---the per-pixel frequency fit is robust against spatial distortions. Consequently, the measured resonance frequency ($f_{88}$) for a given pixel is not intrinsically dependent on the spatial coordinate mapping, though optical imperfections can marginally influence the selection of the region of interest (ROI) determining which pixels contain sufficient signal for analysis.}

\begin{figure*}[htbp]
	\centering
	\includegraphics[width=0.63\columnwidth]{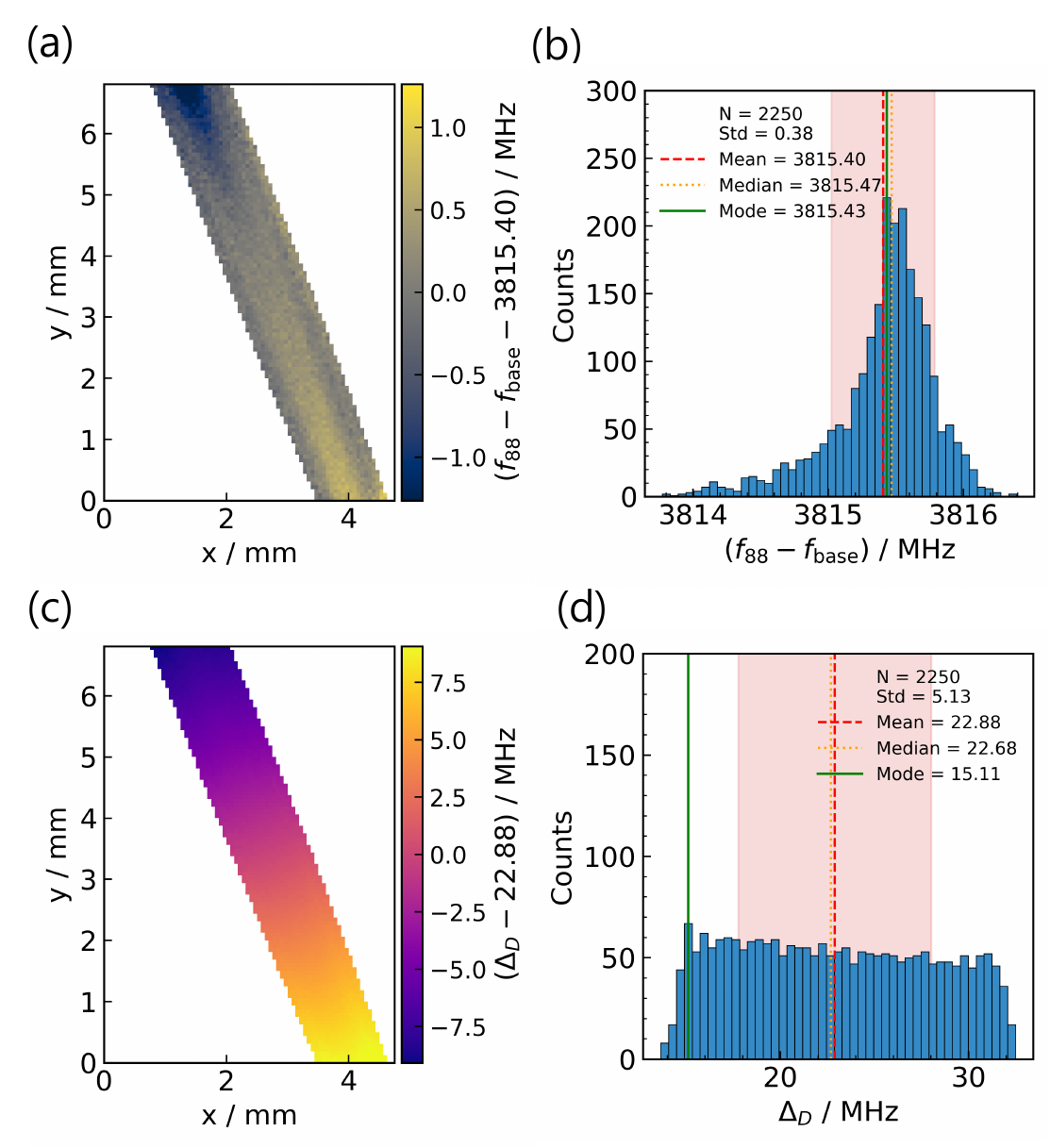}
	\caption{
		Spatial maps and histograms of the fitted spectral parameters obtained from
		the pixel-resolved fluorescence analysis.
		(a) Map of the fitted ${}^{88}$Sr center frequency for the filtered pixel ensemble. 
		(b) Corresponding histogram of the ${}^{88}$Sr center frequencies.
		The vertical lines mark the mean, median, and mode, and the shaded region
		indicates the standard deviation.
		(c) Map of the Doppler shift over the same spatial region.
		(d) Corresponding histogram of the Doppler shifts with the same indicators as in (b).
	}
	\label{fig:spatial_spectroscopy}
\end{figure*}

For each dataset, the valid $f_{88}(x, y)$ values are aggregated into a frequency distribution. 
Fig.~\ref{fig:spatial_spectroscopy}(b) shows a histogram of the filtered pixel ensemble for dataset R1 yielding a mean midpoint frequency of \revred{$f_{\mathrm{base}} + 3815.40$ MHz} with a standard deviation over the accepted image pixels of $0.38 ~\mathrm{MHz}$. 
Note that this wavelength meter (WLM) measurement was referenced to a $778~\mathrm{nm}$ Rb two-photon clock described in section \ref{sec:WLMcalibration}.
We define the resonance frequency of the dataset as the mean of this distribution, with the standard deviation representing the spread. The median and mode are treated as auxiliary descriptors of the distribution symmetry. 

The narrow distribution of measured $f_{88}$ values demonstrates that the Doppler-free resonance frequency estimate is spatially uniform to within ${\sim}1~\mathrm{MHz}$ across the probe–atom interaction region. 
This uniformity can be compared to the Doppler shifts, which exhibit a strong height-dependence as shown in Fig.~\ref{fig:spatial_spectroscopy}(c). 
The corresponding Doppler-shift histogram (Fig.~\ref{fig:spatial_spectroscopy}(d)) confirms that our multi-isotope fit model accurately discriminates the spatially varying in-plane velocity contributions. Because the analysis is image-resolved, transverse velocity components are mapped directly and incorporated into the spatial distribution. 
Hence, we can estimate the expected Doppler spread based on the oven geometry and temperature. 
For our apparatus, the oven nozzle is positioned ${\sim}350~\mathrm{mm}$ from the probe and a mean atomic velocity of $434~\mathrm{m/s}$ is expected at the set oven temperature of \revredd{$315~^{\circ}\mathrm{C}$}. 
The $6~\mathrm{mm}$ spatial extent of our region of interest corresponds to a geometric Doppler spread of ${\sim}16~\mathrm{MHz}$, closely matching our experimental observations.

\revred{To further evaluate possible imaging-related contributions to the extracted resonance frequency, we investigated the operations that can modify the ensemble of fitted spectra, namely the spatial binning and the ROI selection. Repeating the full analysis with $3\times3$ rather than $5\times5$ binning (yielding approximately $2.8\times$ finer effective sampling), while keeping the same physical ROI and filtering criteria, shifted the dataset-averaged $f_{88}$ by less than 0.01~MHz and the mean Doppler shift by approximately 0.02~MHz. Varying the parameters used to determine the adopted pixels for the ROI---which primarily depend on pixel intensity and spatial distance from the principal axis---changed the dataset-averaged $f_{88}$ by less than $\sim$0.03~MHz. Even deliberately loosened ROI selections gave a conservative bound below $\sim$0.1~MHz for the imaging-related ROI-membership sensitivity. Because these residual bounds are significantly smaller than our adopted spatial-reproducibility and model-bias terms, we consider the contribution of a separate imaging-uncertainty term to be negligible.}

\subsection{Velocity estimates of slow atomic beam from a 2-dimensional grating magneto-optical trap}
\label{sec:2DgMOT}

\subsubsection{2D gMOT apparatus}

The 2D gMOT architecture and its performance characteristics are described in comprehensive detail in Ref.~\cite{lee2025}. 
\revred{Briefly, the apparatus shown in Fig.~\ref{fig:apparatus_schematic}(c) utilizes the shared in-vacuum grating chip described previously in section~\ref{sec:spectroscopyMethod}. To form the trap, two grating chips are positioned side-by-side, creating a 2-mm separation gap. Strontium atoms are sourced from an atomic dispenser positioned approximately 5~mm beneath this gap, rather than the remote oven source used in section~\ref{sec:spectroscopyMethod}. To minimize disruptive collisions between the hot thermal atoms and the trapped cold atoms, the atomic jet exiting the dispenser is intentionally tilted by approximately $6^\circ$ relative to the incident cooling beam.
	A two-dimensional quadrupole magnetic field is generated by eight pairs of permanent neodymium magnets. 
	A single 461~nm cooling beam, driving the ${}^{1}S_{0}\rightarrow{}^{1}P_{1}$ transition, is incident on the gratings, creating diffracted beams that overlap with the incident beam to provide two-dimensional optical confinement. Repumping lasers at 679~nm and 707~nm are also applied along the longitudinal axis. 
	The resulting slow atomic beam propagates along the $z$-axis, with its longitudinal velocity distribution determined via an on-resonance "plug-and-probe" time-of-flight (TOF) measurement. The probe beam intersects the atoms downstream to generate a continuous fluorescence signal which is measured with a photomultiplier tube (PMT, Hamamatsu; H14066-01) detector. An additional on-resonance `plug' beam is positioned between the 2D gMOT and the probe beam to selectively block the atomic flux. By abruptly extinguishing the plug beam and measuring the time-delayed arrival of the atoms at the probe region, the velocity of the slow atomic beam can be precisely extracted.
	Notably, the grating chip facilitates the precise characterization of the spatial profile and position of the plug and probe laser beams enabling precise atom velocity measurements.
}

\subsubsection{Modeling atomic velocity and data fitting}
\label{sec:AtomVelcolityModel}
\begin{figure}[!t]
	\centering
	\includegraphics[width=0.9\columnwidth]{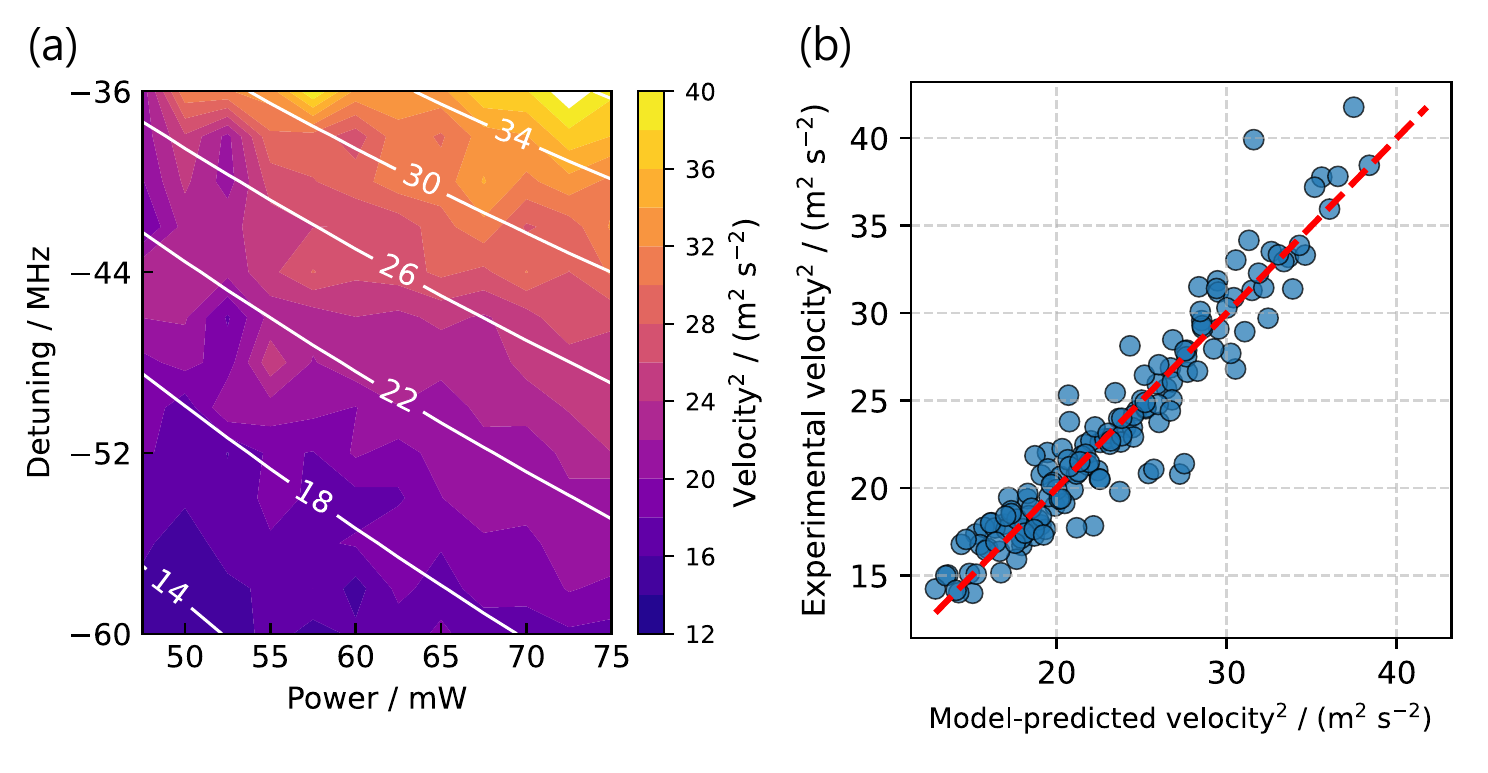}
	\caption{
		Diagnostic plots for the velocity-based fit.
		(a) Measured velocity-squared map as a function of cooling-beam power and detuning, with white contours showing the fitted model.
		(b) Parity plot comparing the model-predicted and experimentally extracted velocity-squared values.
	}
	\label{fig:velocity_fit}
\end{figure}

To extract the absolute resonance frequency, we exploit the functional dependence of the longitudinal velocity of the slow atomic beam on the cooling-laser parameters. To ensure an accurate characterization of the longitudinal velocity distribution, time-of-flight (TOF) measurements are recorded at an optimal distance close to the trap exit. Placing the probe beam farther from the MOT introduces systematic biases, such as velocity-selective losses caused by the finite spatial extent of the probing region. After evaluating several configurations, we optimized the plug-to-probe distance to $6.67(8)~\mathrm{mm}$. This distance provides a balance between maintaining a high signal-to-noise ratio (SNR) for robust fitting and providing a sufficient TOF interval for precise velocity resolution. Fig.~\ref{fig:velocity_fit}(a) shows the experimentally obtained velocity maps of the slow atomic beam from the 2D gMOT as a function of the $461~\mathrm{nm}$ laser intensity and detuning.

The capture velocity $v_c$ of the 2D gMOT is modeled by considering the work done by the radiation pressure over the effective interaction length $D$:
$$
\begin{aligned}
	v_c = \sqrt{\frac{2\hbar k D}{m}\frac{\Gamma}{2}\frac{s}{1 + s + (2\delta/\Gamma)^2}}
\end{aligned}
$$
where $m$ is the atomic mass, $k$ is the cooling-light wavevector, $\Gamma$ is the natural linewidth, $s = I/I_{\mathrm{sat}}$ is the saturation parameter, and $\delta$ is the laser detuning. In this framework, the most probable velocity $v$ of the atoms exiting the trap scales with the square root of the scattering rate, $v \propto \sqrt{\Gamma_{\mathrm{scatt}}}$ \cite{lunden2020}. 
We utilize this model to perform a global least-squares fit on a two-dimensional dataset comprising measured velocities across a range of laser intensities and detunings as shown in Fig.~\ref{fig:velocity_fit}(a). By treating the resonance frequency as a free parameter, we obtain a resonance frequency that is intrinsically linked to the trap dynamics. The resulting fitted velocity map and the associated parity check are summarized in Fig.~\ref{fig:velocity_fit}.

\subsection{Wavemeter Calibration}
\label{sec:WLMcalibration}
The frequency of the $461~\mathrm{nm}$ probe laser is stabilized via a feedback loop referenced to a HighFinesse WS8-10 wavelength meter\cite{couturier2018}. While the manufacturer specifies an absolute accuracy of $10~\mathrm{MHz}$ ($3\sigma$ criterion) under calibrated conditions, this requires routine calibrations which are guarantied for $\sim 1$ hour.
To compensate for long-term drifts, instead of performing frequent recalibrations, we chose to maintain a fixed internal calibration while keeping a record of stable references for post-hoc correction.
Due to intermittent downtime across various optical frequency reference systems, we utilize multiple independent frequency standards as shown in Fig.~\ref{fig:apparatus_schematic}(a) to maintain a traceable calibration record.
According to the measurement records of  our reference lasers, a long-term drift exceeding $60~\mathrm{MHz}$ was observed over the year-long experimental campaign.

For calibrating the WLM readings, we leveraged the optical frequency reference infrastructure at the Korea Research Institute of Standards and Science (KRISS) including a $556~\mathrm{nm}$ Yb ${}^{1}S_{0}\!\rightarrow\!{}^{3}P_{1}$ transition locked laser, a $578~\mathrm{nm}$ Yb optical lattice clock laser and its associated optical frequency comb (OFC), a $759~\mathrm{nm}$ Yb optical lattice laser\cite{Kim2021}, and a $778~\mathrm{nm}$ Rb two-photon optical clock laser.

The $578~\mathrm{nm}$ Yb clock laser, stabilized to an ultra-low expansion (ULE) cavity and referenced in real-time to the OFC, served as our primary calibration anchor. The three retro-beam spectroscopy datasets were corrected using two wavelength-reference schemes: R1 was corrected using the $778~\mathrm{nm}$ Rb two-photon reference, whereas R2 and R3 were corrected using the contemporaneously monitored $759~\mathrm{nm}$ Yb optical lattice laser.
Notably, the $759~\mathrm{nm}$ Yb optical lattice laser is continuously referenced to the WLM for dataset R2-R3 while frequency-locked to the OFC at $394\,798\, 101\, 105\, 918~\mathrm{Hz}$.
The 2D gMOT velocity-based estimate was referenced using the $578~\mathrm{nm}$ laser.

To quantify the WLM measurement accuracy at $461~\mathrm{nm}$ when the calibration is offset by a known amount for a given reference wavelength, we performed a series of controlled tests. 
By introducing synthetic calibration offsets of up to $\pm 300~\mathrm{MHz}$ into the WLM calibration setting for $578~\mathrm{nm}$, we mapped the resulting frequency measurement deviations for three reference lasers: the $578~\mathrm{nm}$ and $759~\mathrm{nm}$ lasers described above, and an additional $556~\mathrm{nm}$ laser.
To stabilize the $556~\mathrm{nm}$ laser, it was frequency-locked to the ${}^{1}S_{0}\!\rightarrow\!{}^{3}P_{1}$ transition of an Yb beam originating from an oven, using a counter-propagating configuration perpendicular to the atomic flux. Given the natural linewidth of $\Gamma = 2\pi \times 182.4 \text{ kHz}$, the resulting frequency reference provides a sufficiently low uncertainty to characterize the performance of the WLM to below $1~\mathrm{MHz}$.

\begin{figure}[!t]
	\centering
	\includegraphics[width=0.55\columnwidth]{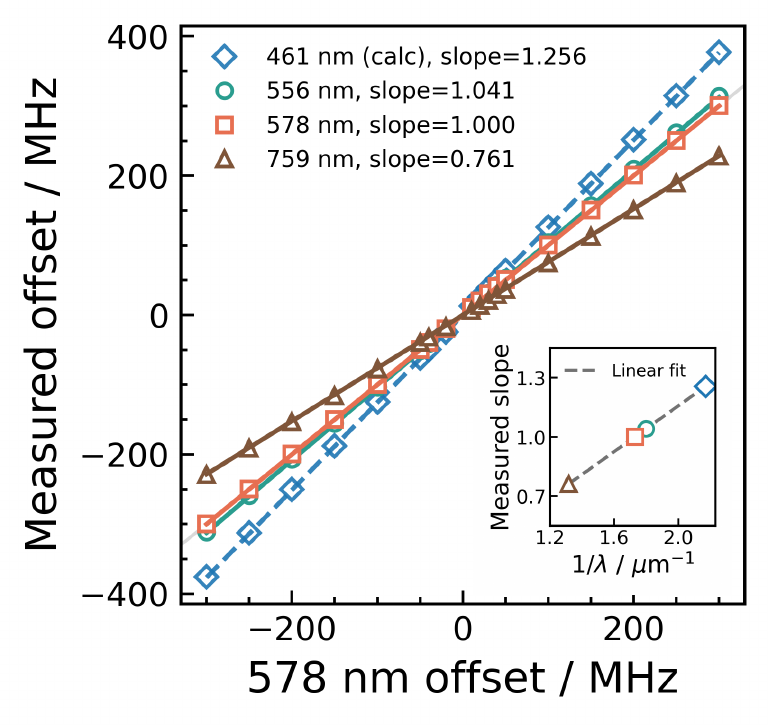}
	\caption{
		Wavelength-meter measurement offset dependence on calibration settings.
		Controlled offsets were applied to a $578~\mathrm{nm}$ calibration laser, and the
		corresponding wavelength-meter shifts were measured at 556, 578, 759, and
		778 nm. The $461~\mathrm{nm}$ response was inferred from the common fractional error.}
	\label{fig:wlm_transfer}
\end{figure}

As shown in Fig.~\ref{fig:wlm_transfer}, these measurements demonstrate that the correction to the WLM measurement is well characterized by a multiplicative frequency-scale model, i.e., the frequency offset ($\delta f$) for a given frequency measurement ($f_1$) is transferred to a different frequency ($f_2$) by the relation $\delta f_1/f_1 = \delta f_2/f_2$.
In this framework, the $461~\mathrm{nm}$ correction can be inferred from a common fractional-error model anchored by the available reference wavelengths.
The measured slopes closely follow the expected optical frequency ratios as shown in the inset of Fig.~\ref{fig:wlm_transfer}. A least squares fit of this data resulted in a fit error of $0.048\%$, which corresponds to negligible values of $5 - 50~\mathrm{kHz}$ for the frequency offsets we observe.
Therefore, we adopt this as a valid frequency transfer model for propagating corrections between the reference wavelengths and $461~\mathrm{nm}$.

\section{Results}
\begin{figure}[!t]
	\centering
	\includegraphics[width=0.55\columnwidth]{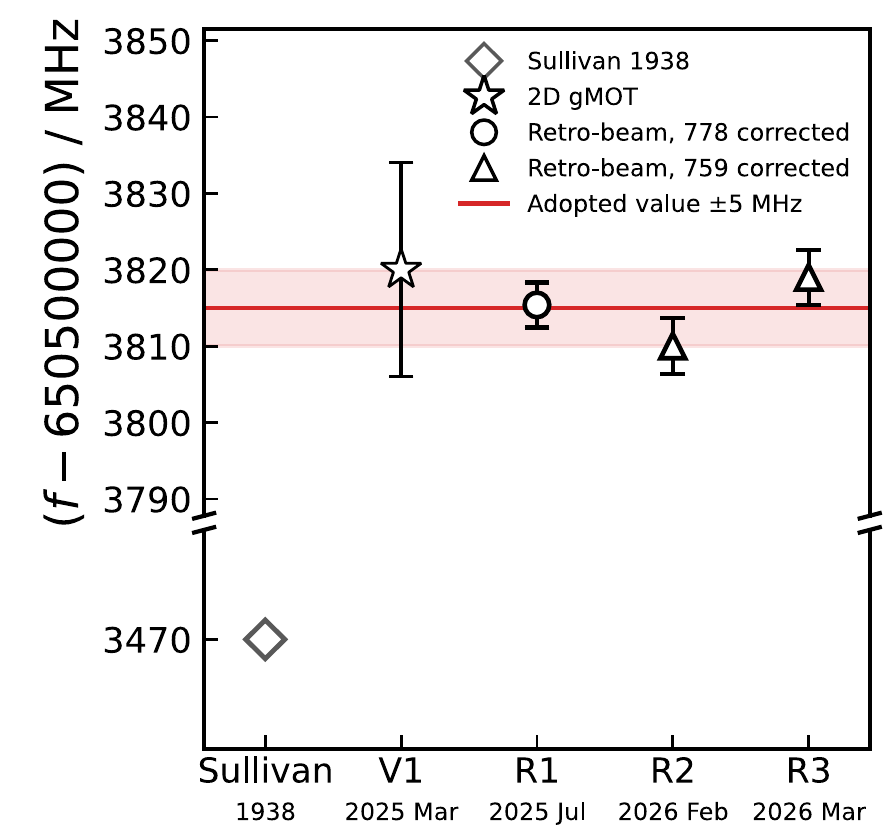}
	\caption{
		Summary of the resonance-frequency estimates for the four datasets V1, R1, R2, and R3.
		Circular and triangular markers denote the spectroscopy results corrected using the 778-nm and 759-nm wavelength references, respectively, while the star denotes the 2D gMOT velocity-based estimate.
		Error bars indicate the total uncertainties for each dataset, obtained by combining the statistical and systematic contributions in quadrature.
		The red horizontal line and shaded band denote the combined resonance frequency and its uncertainty.
		The 1938 solar-spectrum determination is shown for reference\cite{sullivan1938}; its reported uncertainty of \revred{$\pm 310~\mathrm{MHz}$} is omitted for clarity because it exceeds the plotted vertical range.
	}
	\label{fig:resonance_summary}
\end{figure}

Figure~\ref{fig:resonance_summary} summarizes the resonance-frequency estimates obtained from the four datasets V1, R1, R2, and R3 acquired over a 12-month period.
Error bars indicate the total uncertainties for each dataset, obtained by combining the statistical and systematic contributions in quadrature.

\subsection{Statistical uncertainty}

For the spectroscopy datasets, the statistical uncertainty is defined as the standard deviation of the $f_{88}(x,y)$ frequencies within each dataset, such as that shown in Fig.~\ref{fig:spatial_spectroscopy}(b).
The adopted statistical terms are $0.38~\mathrm{MHz}$ for R1, $3.33~\mathrm{MHz}$ for R2, and $3.25~\mathrm{MHz}$ for R3.
The larger values for R2 and R3 mainly reflect the broader spatial distributions associated with the lower signal-to-noise ratio of the corresponding CCD images.
For V1, the statistical uncertainty is determined to be $13.0~\mathrm{MHz}$, which is the fit error of the slow-atom velocity model analysis described in section~\ref{sec:AtomVelcolityModel}.

\subsection{Systematic uncertainty}

The systematic uncertainty budgets for the error bars shown in Fig.~\ref{fig:resonance_summary} are summarized in Table~\ref{tab:run_sys_budget}.
The largest uncertainty comes from the wavelength meter, particularly the manufacturer-specified absolute accuracy of the HighFinesse WS8-10 and the WLM frequency transfer uncertainty.
For R2 and R3, the transfer contribution is $0.56~\mathrm{MHz}$ where the $759~\mathrm{nm}$ reference was monitored contemporaneously with those runs.
For R1, the selected $778~\mathrm{nm}$ reference record was acquired about $13$ days after the spectroscopy run; this temporal separation corresponds to an estimated drift error of up to $\sim 2.4~\mathrm{MHz}$ on the $461~\mathrm{nm}$ measurement.
For the remaining spectroscopy systematic terms, we assign a common value across R1--R3 by adopting the largest shift observed among the three experimental runs.
The reference-laser frequency error is negligible for V1, R2, and R3 on the \revredd{megahertz} scale, but is non-negligible for R1 and contributes significantly to the R1 total systematic uncertainty.
Evaluations of the residual magnetic-field, light-shift, geometric-alignment, and model-bias uncertainties are given in the appendix.
For the V1 dataset, the extracted resonance frequency is scale-invariant with respect to the atomic velocity. Consequently, this determination is inherently robust against systematic uncertainties in the absolute velocity calibration.

\begin{table}[!t]
	\centering
	\small
	\caption{
		Estimated uncertainty budget for each dataset. All entries represent standard uncertainties ($1\sigma$).
	}
	\label{tab:run_sys_budget}
	\setlength{\tabcolsep}{4pt}
	
	\begin{tabular*}{\columnwidth}{@{\extracolsep{\fill}}lccccc@{}}
		\hline
		Source of uncertainty  & \revredd{V1/MHz} & \revredd{R1/MHz} & \revredd{R2/MHz} & \revredd{R3/MHz} \\
		\hline
		Statistical (fit) & 13.0 & 0.38 & 3.33 & 3.25 \\
		Magnetic field (Zeeman) & -- & 0.88 & 0.88 & 0.88 \\
		Light shift & 0.30 & 0.30 & 0.30 & 0.30 \\
		Geometric alignment & -- & 0.44 & 0.44 & 0.44 \\
		Model bias & -- & 0.56 & 0.56 & 0.56 \\
		WLM absolute accuracy & 3.30 & 3.30 & 3.30 & 3.30 \\
		WLM frequency transfer & 4.22 & 2.42 & 0.56 & 0.56 \\
		Reference laser frequency & $<10^{-3}$ & 0.67 & $<10^{-5}$ & $<10^{-5}$ \\
		\hline
		Combined systematic & 5.36 & 4.31 & 3.55 & 3.55 \\
		Total & 14.1 & 4.33 & 4.86 & 4.81 \\
		\hline
	\end{tabular*}
	
\end{table}

The corresponding total uncertainties, obtained by combining the statistical and systematic contributions in quadrature for each dataset, are $14.1~\mathrm{MHz}$ for V1, $4.33~\mathrm{MHz}$ for R1, $4.86~\mathrm{MHz}$ for R2, and $4.81~\mathrm{MHz}$ for R3. Using the four corrected dataset values, we obtain the combined resonance frequency
\revred{
	\[
	\bar{f}_{88} = 650.503\,815(5)~\mathrm{THz}.
	\]
}
The quoted uncertainty includes the manufacturer-specified absolute accuracy of the wavelength meter (HighFinesse WS8-10).
It is evaluated from the statistical and systematic uncertainty budgets of the individual datasets, including correlations among the shared systematic terms, as detailed in the appendix.
The resulting covariance-based uncertainty is $4.45~\mathrm{MHz}$, which is conservatively reported as $5~\mathrm{MHz}$.

As evidenced by the individual budget components, the V1 dataset uncertainty is dominated by statistical fluctuations.
In contrast, the measurements of the R1--R3 datasets reach the systematic limit of our current apparatus, where the dominant uncertainty arises from the wavelength meter, particularly the WS8 absolute-accuracy term and the frequency-transfer uncertainty.
\revred{The agreement between the two independent methods provides strong confidence in the accuracy of the final determined value.}

\section{Summary}
We have determined the absolute frequency of the ${}^{1}S_{0}\!\rightarrow\!{}^{1}P_{1}$ transition using two independent methods facilitated by an in-vacuum photonic grating chip. 
This chip-defined geometry provides superior mechanical stability and enables precise control over the atom-light interaction region. 
By leveraging spatially resolved 2D fluorescence spectroscopy, we have reconstructed high-fidelity maps of the Doppler-shifted resonances across the interaction volume via  multi-isotope fit analysis. 
In addition to our spectroscopic measurements, we utilized the velocity-mapped cooling dynamics of the 2D gMOT to provide an independent determination of the resonance frequency. 
This approach is grounded in a first-principles analytical model of the relationship between trap parameters and atomic velocity.
The robustness of both methods for this determination is validated by the consistency between the four datasets and ensemble statistics derived from thousands of independent pixel spectra. Our measurement corresponds to a 345 MHz shift relative to the center value reported in the 1938 solar-spectrum analysis.
Notably, while our estimate deviates from the reported \revred{310 MHz} standard uncertainty (1$\sigma$), it remains within the expanded uncertainty (2$\sigma$) of the 1938 estimate.
This work provides a substantial improvement in precision, reducing the uncertainty by more than a factor of 50.
By providing a high-accuracy, high-precision estimate of the ${}^{1}S_{0}\!\rightarrow\!{}^{1}P_{1}$ transition frequency, this work establishes a metrological foundation for advanced Sr-based quantum systems directed towards precision spectroscopy, optical frequency synthesized references, and quantum metrology.

\section{Appendix}
\subsection{Common systematic uncertainty sources for all datasets}
There are two sources of systematic uncertainty that are common to all datasets: light-shift and WLM absolute accuracy. Details of the evaluated uncertainty values are described below.

\subsubsection{Light-shift uncertainty}
\label{sec:appendix_LS}
A full theoretical evaluation of the light shift for the ${}^{1}S_{0}\!\rightarrow\!{}^{1}P_{1}$ transition is not carried out in this study because reliable decay-rate information is not available for several relevant excited states. Notably, part of the intensity variation is already incorporated into the spatially resolved analysis because different pixel spectra correspond to different local probe intensities, indicating that the light-shift uncertainty ($u_{\mathrm{LS}}$) cannot be more than the statistical scatter of $0.38~\mathrm{MHz}$.

As an auxiliary check on probe-power dependence, we compared three additional runs measured at probe powers of $170.5~\mu\mathrm{W}$, $365.5~\mu\mathrm{W}$ and $555.5~\mu\mathrm{W}$. The corresponding resonance frequencies are $650\,503\,815.24~\mathrm{MHz}$, $650\,503\,815.40~\mathrm{MHz}$ and $650\,503\,814.82~\mathrm{MHz}$, giving a mean of $650\,503\,815.15~\mathrm{MHz}$ and a sample standard deviation of $0.30~\mathrm{MHz}$. The actual light-shift uncertainty is likely smaller than this spread; however, as a conservative upper bound we adopt the common light-shift uncertainty for all datasets as
\[
u_{\mathrm{LS}}=0.3~\mathrm{MHz}.
\]

\subsubsection{WLM absolute accuracy uncertainty}
The manufacturer-specified $1\sigma$ accuracy of the HighFinesse WS8 when calibrated is $3.3~\mathrm{MHz}$ ($10~\mathrm{MHz}$ according to $3\sigma$ criterion for $375 – 800~\mathrm{nm}$). This common systematic uncertainty is included to all datasets as
\[
u_{\mathrm{wlm}}=3.3~\mathrm{MHz}.
\]

\subsection{Uncertainty budget for the datasets R1, R2, R3}

This section summarizes the adopted uncertainty budget for the three spatially resolved spectroscopy determinations, R1--R3, shown in Fig.~\ref{fig:resonance_summary}. The statistical and systematic contributions are evaluated as described below.

\subsubsection{Statistical uncertainty}

\revred{For each spectroscopy dataset, we use the standard deviation of the fitted ${}^{88}$Sr center frequencies over the accepted pixels as a conservative within-image reproducibility term.  While the standard error of the mean ($\sigma/\sqrt{N}$) is typically used to represent the uncertainty of $N$ independent measurements, the pixel spectra in our spatially resolved analysis are not strictly independent repeated measurements of the same quantity. Although all pixel spectra share the same probe-laser frequencies, each pixel corresponds to a different local probe intensity, and neighboring pixels can be correlated through the imaging point-spread function. Consistent with this lack of strict independence, the observed frequency distribution exhibits a skewness indicating a departure from a purely Gaussian, independent-sample model. Rather than attempting to isolate pure shot noise in the presence of these systematic spatial biases, we adopt the standard deviation of the distribution as a conservative spatial-reproducibility term to define the statistical uncertainty.
}
The adopted statistical uncertainties for each dataset are 
\[
u_{\mathrm{stat,R1}} = 0.38~\mathrm{MHz},\qquad 
u_{\mathrm{stat,R2}} = 3.33~\mathrm{MHz},\qquad
u_{\mathrm{stat,R3}} = 3.25~\mathrm{MHz}.
\]
\begin{figure}[!t]
	\centering
	\includegraphics[width=0.8\columnwidth]{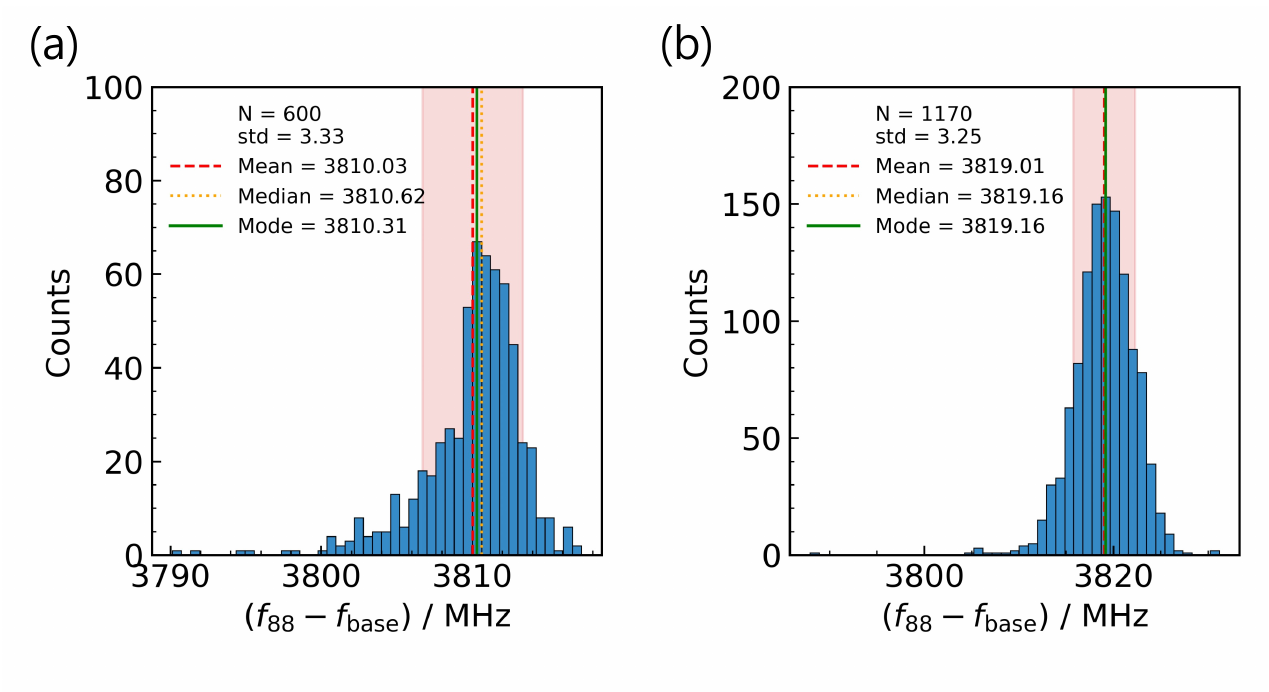}
	\caption{
		Histograms of the fitted ${}^{88}$Sr resonance values from the accepted pixel spectra for datasets R2 (a) and R3 (b). The statistical uncertainty is taken as the standard deviation of each distribution, giving $3.33~\mathrm{MHz}$ for R2 and $3.25~\mathrm{MHz}$ for R3. The vertical lines mark the mean, median, and mode.
	}
	\label{fig:appendix_R2R3}
\end{figure}

\subsubsection{Geometric alignment uncertainty}
\label{sec:GeomAlignUncert}
The geometric alignment uncertainty ($u_{\mathrm{geom}}$) refers to the residual frequency shift arising from imperfect retro-reflection of the probe beam when the atomic beam trajectory is not exactly perpendicular to the probe-beam propagation direction. This uncertainty contribution appears when the return beam deviates from exact antiparallel propagation by an angle $\delta\theta_{\rm geom}$. The midpoint of the counter-propagating resonances then acquires a residual first-order Doppler contribution,
\[
u_{\rm geom}\approx \frac{v}{2\lambda}\,\delta\theta_{\rm geom} \approx 0.44~\mathrm{MHz},
\]
where $v$ is the mean atomic-beam velocity and $\lambda$ is the probe wavelength.

In practice, the retro-reflection is aligned by back-coupling the return beam through an iris placed at an optical propagation distance of approximately $700~\mathrm{mm}$ from the grating chip.
Taking the transverse mismatch to be comparable to the measured $1/e^2$ beam radius of $0.66~\mathrm{mm}$ gives $\delta\theta_{\rm geom}\simeq 0.0540^\circ.$
Using $v=434~\mathrm{m/s}$ and $\lambda=461~\mathrm{nm}$ yields a common geometric alignment uncertainty of $0.44~\mathrm{MHz}$ for the datasets R1--R3.

\subsubsection{WLM frequency transfer uncertainties}

The uncertainty in frequency measurements at $461~\mathrm{nm}$ due to the WLM ($u_{\mathrm{trans}}$) depends on multiple components involved in the reference-transfer chain for the frequency measurement. We first established the intrinsic stability of the WLM by monitoring a $578~\mathrm{nm}$ clock laser that was locked to a high-finesse ultra-low expansion (ULE) cavity.
The $578~\mathrm{nm}$ laser is beat against an OFC and recorded using a K+K frequency counter to measure the absolute frequency in real-time. Over a typical experimental timescale of $10,000\text{ s}$, the laser exhibits a frequency instability of $\pm 17.04\text{ Hz}$. This stabilized laser was employed as a reference to calibrate the WLM, after which its stability was characterized via Allan deviation analysis. The fractional frequency instability of the WLM was found to be $1.5 \times 10^{-10}$ at $1\text{ s}$, maintaining a level below $1.5 \times 10^{-10}$ up to $10,000\text{ s}$. This corresponds to an intrinsic WLM frequency uncertainty of approximately $77.7\text{ kHz}$, which is sufficiently narrow to have a negligible impact on the transferred WLM measurement uncertainties in the order of $1~\mathrm{MHz}$.

The dataset R1 was referenced to an optical frequency standard based on the $^{87}\mathrm{Rb}$ $5S_{1/2}~(F=2) \rightarrow 5D_{5/2}~(F'=4)$ two-photon transition at $778~\mathrm{nm}$. Comparison of the WLM records with the CIPM recommended frequency of $385\,285\,566\,370.4~\mathrm{kHz}$ revealed a mean offset of $-13.02~\mathrm{MHz}$ and a standard deviation of $0.20~\mathrm{MHz}$. These parameters were applied as systematic corrections to the measured $461~\mathrm{nm}$ wavelengths. Assuming a proportional transfer model for the WLM scaling, the statistical fluctuation at $778~\mathrm{nm}$ translates to a frequency uncertainty of $0.34~\mathrm{MHz}$ when measuring $461~\mathrm{nm}$ wavelengths.

An additional term is included for R1 because the selected $778~\mathrm{nm}$ reference was recorded 13 days after acquiring the spectroscopy dataset. Based on periodic reference frequency monitoring throughout the year-long campaign, we established a long-term WLM drift rate of $5.58~\mathrm{MHz/month}$ at $461~\mathrm{nm}$. This long-term drift uncertainty contributes an additional $2.4~\mathrm{MHz}$ to the WLM measurement uncertainty for this dataset.

Combining the transferred $778~\mathrm{nm}$ scatter and the 13-day drift error in quadrature gives an R1 dataset WLM frequency transfer uncertainty of 
\[
u_{\mathrm{trans,R1}}=2.42~\mathrm{MHz}.
\]

\begin{figure}[!t]
	\centering
	\includegraphics[width=0.45\columnwidth]{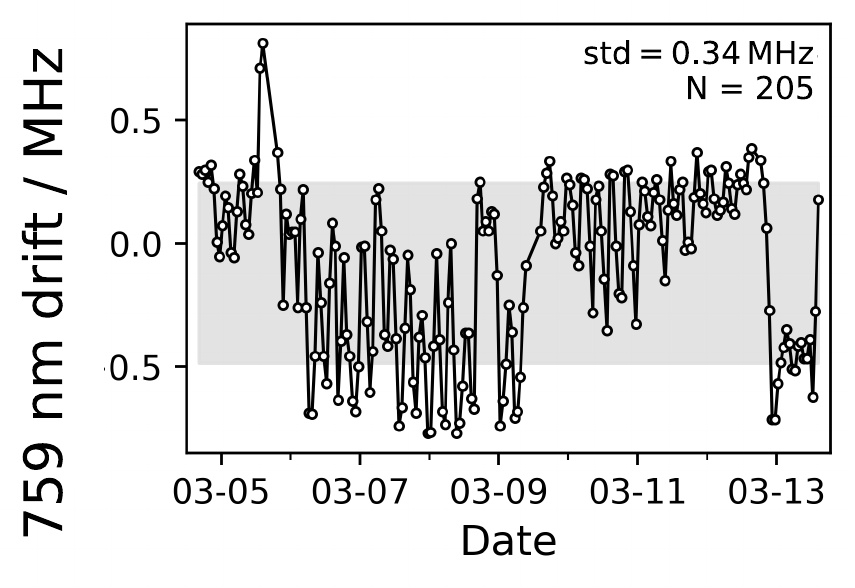}
	\caption{
		\revredd{Week-long $759~\mathrm{nm}$  WLM stability record monitored contemporaneously with the R2 and R3 runs. The shaded band denotes $\pm1\sigma$.}
	}
	\label{fig:appendix_778_759}
\end{figure}

Datasets R2 and R3 are referenced to the week-long $759~\mathrm{nm}$ WLM stability record shown in Fig.~\ref{fig:appendix_778_759}. Because this record was monitored contemporaneously with the spectroscopy runs, no additional long-term drift component is required. The observed standard deviation is $0.34~\mathrm{MHz}$ at $759~\mathrm{nm}$, corresponding to $0.56~\mathrm{MHz}$ when measuring $461~\mathrm{nm}$. Thus, the shared transfer uncertainty for the R2 and R3 datasets is
\[
u_{\mathrm{trans,R2}} = u_{\mathrm{trans,R3}} = 0.56~\mathrm{MHz}.
\]

\subsubsection{Residual magnetic-field uncertainty}

The residual magnetic-field contribution to the uncertainty is carried as a common term for datasets R1--R3. A conservative value assigned from the measured fluxgate magnetometer readings of magnitudes up to $\sim 630~\mathrm{mG}$ in the vicinity of the spectroscopy region is evaluated to be
\[
u_{\mathrm{B}} = 0.88~\mathrm{MHz}.
\]

\subsubsection{Model bias}

The model-dependent systematic uncertainty ($u_{\mathrm{model}}$) is comprised of three primary components. First, we evaluated the sensitivity of the extracted resonance frequency to the spatial inhomogeneity of the $^{87}\mathrm{Sr}$ hyperfine component distribution within the interaction region. By comparing spatially invariant hyperfine weight models against a linearly varying model, this contribution was determined to be $0.32~\mathrm{MHz}$. Second, a line-shape sensitivity term of $0.25~\mathrm{MHz}$ was assigned to account for the discrepancies observed for the resonance frequency when fitting with Lorentzian versus pseudo-Voigt profiles. Finally, an uncertainty of $0.38~\mathrm{MHz}$ was established for the mean Doppler-shift correction, derived from a sensitivity analysis of the model fit function under varying parameters.
Combining the three model-bias contributions in quadrature gives a common systematic term for R1--R3 of
\[
u_{\mathrm{model}} = 0.56~\mathrm{MHz}.
\]

\subsubsection{Reference laser frequency}

The reference-frequency uncertainty for dataset R1 was assigned a value of $u_{\mathrm{ref,R1}} = 0.67~\mathrm{MHz}$, corresponding to the natural linewidth of the $778~\mathrm{nm}$ Rb two-photon transition. In contrast, for datasets R2 and R3, the frequency reference was a $759~\mathrm{nm}$ Yb optical lattice laser stabilized to the OFC. The frequency uncertainty for the $759~\mathrm{nm}$ laser is governed by the stability of the hydrogen maser used to reference the OFC. We estimate this contribution to be $u_{\mathrm{ref,R2}} = u_{\mathrm{ref,R3}} = 1.54~\mathrm{Hz}$, which was obtained from Allan deviation measurements of the hydrogen maser. This value is negligible relative to the megahertz-scale systematic uncertainties.

\subsection{Uncertainty budget for the dataset V1}
\label{sec:appendix_v1_budget}

The velocity-based determination V1 is obtained from the independent 2D gMOT analysis described in the main text. The related uncertainty budget is treated separately from the R1--R3 datasets because the resonance frequency is inferred from the atomic velocity analysis rather than from spatially resolved fluorescence spectra.

\subsubsection{Statistical uncertainty}

The statistical uncertainty for dataset V1 is taken from the fit error of the resonance frequency associated with the 2D gMOT-based velocity measurements according to the model described in section~\ref{sec:2DgMOT}. The fidelity of the fit is reflected in the parity plot shown in Fig.~\ref{fig:velocity_fit}(b). The adopted value is
\[
u_{\mathrm{stat,V1}}=13.0~\mathrm{MHz}.
\]

\subsubsection{Wavelength-transfer uncertainty}

The primary systematic uncertainty for the dataset V1 is attributed to the long-term drift rate of the WLM ($5.58~\mathrm{MHz/month}$), evaluated based on the time elapsed between the reference frequency measurement and the acquisition of the V1 dataset.
The adopted value is
\[
u_{\mathrm{trans,V1}}=4.22~\mathrm{MHz}.
\]

\subsection{Summary of adopted uncertainties}

The adopted total uncertainties are $14.06~\mathrm{MHz}$ for V1, $4.33~\mathrm{MHz}$ for R1, $4.86~\mathrm{MHz}$ for R2, and $4.81~\mathrm{MHz}$ for R3, as summarized in Table~\ref{tab:run_sys_budget}. The systematic uncertainty budgets for spectroscopy datasets R1--R3 are comprised of common-mode contributions, with the exception of those originating from the WLM transfer chain and the respective reference lasers. The uncertainty for dataset V1 is dominated by the statistical variance of the velocity-distribution analysis. These uncertainty components are integrated into the combined uncertainty analysis detailed in the following section.

\subsection{Combined resonance frequency value and uncertainty}
\label{sec:appendix_combined_value}

The combined resonance frequency value used in the manuscript is taken as the weighted mean of the four frequency corrected datasets listed below. 
\begin{table}[!ht]
	\centering
	\small
	\caption{$f_{88}$ estimates in units of \revredd{megahertz}} 
	\label{tab:combinedResFreq}
	\setlength{\tabcolsep}{4pt}
	\begin{tabular*}{\columnwidth}{@{\extracolsep{\fill}}ccccc@{}}
		\hline
		V1 & R1 & R2 & R3 & Combined \\
		\hline
		650\,503\,820.00 & 650\,503\,815.40 & 650\,503\,810.03 & 650\,503\,819.01 & 650\,503\,815.10 \\
		\hline
	\end{tabular*}
\end{table}

The uncertainty of this weighted average is evaluated from the covariance matrix
$\mathbf{C}=\mathbf{C}_{\mathrm{diag}}+\mathbf{C}_{\mathrm{off}}$,
where \(\mathbf{C}_{\mathrm{diag}}\) contains the diagonal variance terms and \(\mathbf{C}_{\mathrm{off}}\) contains the correlated systematic terms in the off-diagonal matrix elements.

The contributions common to all four determinations are the light-shift term ($u_{\mathrm{LS}}$) and the manufacturer-specified WS8 absolute accuracy term ($u_{\mathrm{wlm}}$) so that
\[
C_{\mathrm{off,1}}=u_{\mathrm{LS}}^2+u_{\mathrm{wlm}}^2=10.98~\mathrm{MHz}^2.
\]

For the datasets R1--R3, additional off-diagonal (correlated) terms are included for the covariance matrix element
\[
C_{\mathrm{off,2}}
=
C_{\mathrm{off,1}}
+
u_B^2
+
u_{\mathrm{model}}^2
+
u_{\mathrm{geom}}^2
=
12.26~\mathrm{MHz}^2,
\]
and finally, the reference laser scheme was identical for the R2 and R3 datasets giving another element
\[
C_{\mathrm{off,3}}
=
C_{\mathrm{off,2}} + u_{\mathrm{trans,R2-R3}}^2
=
12.58~\mathrm{MHz}^2.
\]

The remaining diagonal terms are
\[
C_{\mathrm{diag,V1}}=u_{\mathrm{stat,V1}}^2+u_{\mathrm{trans,V1}}^2+C_{\mathrm{off,1}}
\qquad
C_{\mathrm{diag,R1}}=u_{\mathrm{stat,R1}}^2+u_{\mathrm{trans,R1}}^2+u_{\mathrm{ref,R1}}^2+C_{\mathrm{off,2}},
\]
\[
C_{\mathrm{diag,R2}}=u_{\mathrm{stat,R2}}^2+C_{\mathrm{off,3}},
\qquad
C_{\mathrm{diag,R3}}=u_{\mathrm{stat,R3}}^2+C_{\mathrm{off,3}}.
\]

\[
\mathbf{C}
=
\begin{pmatrix}
	C_{\mathrm{diag,V1}} & C_{\mathrm{off,1}} & C_{\mathrm{off,1}} & C_{\mathrm{off,1}} \\
	C_{\mathrm{off,1}} & C_{\mathrm{diag,R1}} & C_{\mathrm{off,2}}& C_{\mathrm{off,2}} \\
	C_{\mathrm{off,1}} & C_{\mathrm{off,2}} & C_{\mathrm{diag,R2}} & C_{\mathrm{off,3}} \\
	C_{\mathrm{off,1}} & C_{\mathrm{off,2}} & C_{\mathrm{off,3}} & C_{\mathrm{diag,R3}}
\end{pmatrix}
=
\begin{pmatrix}
	197.79 & 10.98 & 10.98 & 10.98 \\
	10.98 & 18.73 & 12.26 & 12.26 \\
	10.98 & 12.26 & 23.66 & 12.58 \\
	10.98 & 12.26 & 12.58 & 23.14
\end{pmatrix}
\ \mathrm{MHz}^2
\]
where the \revredd{hertz}-level reference-laser terms for V1, R2 and R3 are negligible at this precision.

For the weighted average
\[
\bar f_{88}
=
\mathbf{w}^{\mathsf T}\mathbf{f}_{88}
=
650\,503\,815.10~\mathrm{MHz},
\]
where $\mathbf{w}$ is the normalized weight vector, and the uncertainty of the weighted mean is 

\[
u_{\mathrm{w}}
=
\revred{\sqrt{\mathbf{w}^{\mathsf T} \mathbf{C} \, \mathbf{w}}
	=}
(\mathbf{1}^{\mathsf{T}} \mathbf{C}^{-1} \mathbf{1})^{-1/2}
=
3.90~\mathrm{MHz}.
\]

The Birge ratio is calculated to be $R_{\mathrm{B}}=\sqrt{\chi^2/\nu}=1.14$, where $\chi^2$ is the generalized Chi-squared statistic and the degree of freedom is $\nu = 3$.
Since \(R_{\mathrm{B}}>1\), the covariance-based combined uncertainty is
\[
u_{\mathrm{total}}=R_{\mathrm{B}}\,u_{\mathrm{w}}=4.45~\mathrm{MHz}.
\]

For the quoted resonance frequency value in the main text, we conservatively round the combined uncertainty to $5~\mathrm{MHz}$ and report
\revred{
	\[
	\bar{f}_{88}=650.503\,815(5)~\mathrm{THz}.
	\]
}

\ack{This work was supported by Korea Research Institute for defense Technology planning and advancement (KRIT), funded by Defense Acquisition Program Administration (DAPA) (KRIT-CT-23-001).
The authors would like to thank S. L. Lee and J. J. Choi for their assistance with characterizing the wavelength meter.
}

\data{The data that support the findings of this study are available from the corresponding author upon reasonable request.
}

\bibliographystyle{iopart-num}
\bibliography{refs}

\end{document}